\documentclass[10pt,twocolumn,letterpaper]{article}

\usepackage{cvpr}              

\usepackage{verbatim}

\usepackage{tikz}
\usepackage{tikz-3dplot}
\usetikzlibrary{arrows.meta}
\usepackage[table]{xcolor}
\definecolor{bestcol}{RGB}{198,239,206}     
\definecolor{secondcol}{RGB}{189,215,238}  
\usepackage{multirow}

\usepackage{amsthm}

\usepackage[dvipsnames]{xcolor}
\definecolor{lightred}{RGB}{255, 153, 153}

\newcommand{\TODO}[1]{\textbf{\color{red}[TODO: #1]}}
\newcommand{\crv}[1]{{\color{lightred}#1}}

\newcommand{\mypar}[1]{\vspace{4pt}\noindent\textbf{#1}~}

\renewcommand{\TODO}[1]{}

\renewcommand{\crv}[1]{#1}

\newcommand{\site}{\mathbf{s}}
\newcommand{\vertex}{\mathbf{v}}
\newcommand{\centroid}{\mathbf{c}}
\newcommand{\point}{\mathbf{p}}
\newcommand{\sdf}{\phi}
\newcommand{\eigenvector}{\mathbf{e}}
\newcommand{\normal}{\mathbf{n}}
\newcommand{\targetpt}{\mathbf{p}}
\newcommand{\score}{\mathcal{S}}
\newcommand{\voronoicell}{\mathcal{V}}
\newcommand{\tetravol}{V}

\newcommand{\tetra}{\mathcal{T}}

\newcommand{\adjtetraset}{\mathcal{D}}

\usepackage{graphicx,scalerel}
\newcommand\wh[1]{\hstretch{2}{\hat{\hstretch{.5}{#1}}}}

\definecolor{cvprblue}{rgb}{0.21,0.49,0.74}
\usepackage[pagebackref,breaklinks,colorlinks,citecolor=cvprblue]{hyperref}

\def\paperID{384} 
\def\confName{3DV\xspace}
\def\confYear{2026\xspace}

\title{DCCVT: Differentiable Clipped Centroidal Voronoi Tessellation}

\author{
    Wylliam Cantin Charawi$^{1}$ \quad 
    Adrien Gruson$^{1}$ \quad 
    Jane Wu$^{2}$ \quad 
    Christian Desrosiers$^{1}$ \quad 
    Diego Thomas$^{3}$ \\
    $^{1}$École de Technologie Supérieure \quad 
    $^{2}$UC Berkeley \quad 
    $^{3}$Kyushu University \\
}

\begin{document}
\maketitle

\begin{abstract}
\vspace{-4mm}
While Marching Cubes (MC) and Marching Tetrahedra (MTet) are widely adopted in 3D reconstruction pipelines due to their simplicity and efficiency, their differentiable variants remain suboptimal for mesh extraction. This often limits the quality of 3D meshes reconstructed from point clouds or images in learning-based frameworks. In contrast, clipped CVTs offer stronger theoretical guarantees and yield higher-quality meshes. However, the lack of a differentiable formulation has prevented their integration into modern machine learning pipelines. To bridge this gap, we propose DCCVT, a differentiable algorithm that extracts high-quality 3D meshes from noisy signed distance fields (SDFs) using clipped CVTs. We derive a fully differentiable formulation for computing clipped CVTs and demonstrate its integration with deep learning-based SDF estimation to reconstruct accurate 3D meshes from input point clouds. Our experiments with synthetic data demonstrate the superior ability of DCCVT against state-of-the-art methods in mesh quality and reconstruction fidelity. \crv{https://wylliamcantincharawi.dev/DCCVT.github.io/}
\end{abstract}

\vspace{-4mm}
\section{Introduction}
\vspace{-1mm}

\label{sec:intro}

High-quality three-dimensional content is essential across a wide range of fields, including AR/VR, digital simulation, healthcare, gaming, and filmmaking. Among the various 3D representations, discrete formats like 3D meshes are widely used due to their simplicity in rendering and manipulation. However, generating accurate 3D content from real-world data, such as 2D images or partial or noisy point clouds, is a highly challenging task that typically demands expert knowledge and considerable manual effort. As a result, extensive research has focused on assisting or fully automating this process. One common approach involves using implicit representations, such as Signed Distance Fields (SDFs), which are optimized from observations and later converted into discrete 3D meshes through a post-processing step. This method is effective because smooth SDFs are well-suited for optimization techniques.

\begin{figure}[t]
    \centering 
    \includegraphics[width=1\columnwidth]{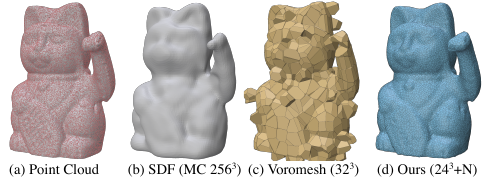}
    \vspace{-6mm}
    \caption{Comparison of 3D mesh reconstruction from an unoriented point cloud. (a) Input point cloud. (b) Input Signed Distance Field (SDF) reconstructed using Marching Cubes at \( 256^3 \) resolution. (c) Voromesh baseline with \( 32^3 \) sites. (d) Our proposed DCCVT method with \(24^3 \) sites plus near sampling sites, achieving more accurate and regular surface reconstruction. }
    \label{fig:teaser}
    \vspace{-5mm}
\end{figure}

In recent years, significant attention has been devoted to optimizing and regularizing SDFs from multi-view images or 3D point clouds. However, the final step of extracting a 3D mesh is typically handled by simple methods that are decoupled from the optimization process, often resulting in sub-optimal reconstructions. To address this, differentiable versions of Marching Cubes (DMC) \cite{liao2018dmc} and Marching Tetrahedra (DMTet) \cite{shen2021dmt} have been introduced, enabling the integration of mesh extraction directly into the optimization loop. These methods have shown improved mesh reconstruction quality when using SDFs or point clouds as input. Nevertheless, we argue that the underlying algorithms, Marching Cubes and Marching Tetrahedra, are themselves not optimal for high-quality mesh extraction, and thus their differentiable counterparts (DMC and DMTet) remain fundamentally limited.

Marching Cubes operates on a fixed grid where SDF values are sampled at the grid's vertices, and a triangular mesh is extracted along edges where the SDF crosses zero. While this method is fast and efficient, it often suffers from poor tessellation due to the fixed orientation of grid cells, leading to irregular triangles and approximation errors, particularly for complex geometries. As a result, post-processing is typically required to produce a more regular 3D mesh. On the other hand, Marching Tetrahedra offers greater flexibility by using Delaunay tetrahedralization, which allows the tessellation to better adapt to the shape of the implicit surface. However, irregularities in the mesh remain challenging to eliminate completely. Recent studies have shown that Centroidal Voronoi Tessellations (CVT) can produce volumetric and surface discretizations with significantly improved regularity \crv{\cite{Du2003tetmeshcvt}}. In addition, CVTs provide provably more accurate approximations of implicit shapes compared to both Marching Cubes and Marching Tetrahedra.

In this paper, we propose DCCVT, a differentiable algorithm for the clipped CVT (CCVT) algorithm~\cite{wang2016clippedCVT}. CCVT has proven advantages compared to MC and MTet and we demonstrate that DCCVT also has significant advantages compared to DMC and DMTet. While our primary focus is on 3D mesh reconstruction from input point clouds, our core contribution of differentiable clipped CVT also has broader applicability to other 3D mesh reconstruction pipelines. Our contributions are:
\begin{itemize}
\item A Centroidal Voronoi Tesselation extraction from a sparse point cloud;
\item An expressive end-to-end optimization of Voronoi site position and their associated SDF values through a clipping operation;
\item An iterative upsampling method improving the overall robustness of our method.
\end{itemize}

\vspace{-1mm}
\section{Related Work}
\vspace{-1mm}

\mypar{Explicit methods.}

Explicit methods directly connect sample points to form a mesh. A representative example is the Ball Pivoting Algorithm (BPA) introduced by \citet{bernardini2002ball}, which forms a triangle whenever a ball of a user-specified radius can touch three points without enclosing any others. This simple heuristic yields watertight meshes and remains widely used in practice. However, BPA suffers from notable limitations, including sensitivity to noise, non-uniform sampling densities, and the need to tune a single ball radius that may not suit all surface regions.
In contrast, Poisson Surface Reconstruction \cite{kazhdan2006poisson,kazhdan2013poissonscreened} formulates mesh recovery as solving a well-conditioned sparse linear system using local basis functions. It does not rely on radius heuristics and is generally more robust to noise, but requires oriented point clouds, which are often unavailable in raw scans.
Normal estimation algorithms \cite{guerrero2018pcpnet,ben2019nesti,ben2020deepfit,lenssen2020deep} could be used to estimate point sample orientations, but existing methods still yield noisy predictions.

Deep learning approaches to point set triangulation have also been proposed \cite{luo2021deepdt,sharp2020pointtrinet,son2024dmesh,zhang2025learning,zhang2023dmnet,maruani2024ponq}.
DeepDT \cite{luo2021deepdt} reconstructs surfaces by predicting in/out labels of tetrahedrons directly from the point cloud and its corresponding Delaunay triangulation.
PointTriNet \cite{sharp2020pointtrinet} proposes candidate triangles and another classifies whether each triangle should appear in the final triangulation.
DMNet \cite{zhang2023dmnet} and followup work in \citet{zhang2025learning} models the Delaunay triangulation as a dual graph and embeds local geometric features into its structure.

\mypar{Implicit methods.}
A number of learning-based approaches to surface reconstruction involve predicting a continuous SDF of 3D surfaces directly from unstructured point sets \cite{park2019deepsdf,jiang2020sdfdiff,erler2020points2surf,gropp2020implicit,ma2020neural,sitzmann2020implicit,chou2022gensdf,wang2023neural,chen2024neuraltps,li2024implicit,zhang2025high,zimo2025hotspot}.
One of the earliest works in this area was DeepSDF \cite{park2019deepsdf}, which introduced a latent‑conditioned neural network that learns a continuous SDF for an entire shape class, enabling high‐quality shape representation, interpolation, and completion from partial/noisy data.
Points2Surf \cite{erler2020points2surf} presented a patch‐based framework that learns an implicit surface from raw point clouds without normals by combining detailed local patches with coarse global SDF sign predictions; this yields more accurate reconstructions and better generalization to unseen shapes.
NeuralTPS \cite{chen2024neuraltps} infers an SDF from a single sparse point cloud without any learned shape priors or normal inputs by leveraging thin‐plate‐spline surface parameterizations to generate coarse surface samples during training.
\citet{zhang2025high} proposes a two-stage approach that combines SDF learning with adaptive Delaunay meshing algorithm.
The recent Hotspot method \cite{zimo2025hotspot} can infer a neural SDF within minutes.
Besides SDF representations, a number of neural implicit representations have also been proposed for surface reconstruction from point clouds \cite{tatarchenko2017octree,mescheder2019occupancy,jiang2020local,peng2020convolutional,tancik2020fourier,lipman2021phase,williams2021neural,yifan2021geometry,ben2022digs,shen2024spacemesh}.

While implicit representations can capture smooth and accurate surfaces, detail preservation often suffers when applying standard mesh extraction methods such as Marching Cubes or Marching Tetrahedra. Consequently, current research is focusing on integrating mesh extraction into the learning process to preserve fidelity, with differentiable extraction emerging as a key approach.

\mypar{Differentiable mesh extraction.}
\citet{liao2018dmc} introduced Deep Marching Cubes, a differentiable variant of Marching Cubes, enabling end-to-end training from point clouds. However, the approach is limited by its requirement for an axis-aligned voxel grid, which constrains detail resolution. To address this, DMTet \cite{shen2021dmt} replaces the grid with a tetrahedral discretization, removing axis alignment constraints and enabling finer reconstructions. Similarly, \citet{wu2025sparse3D} proposed a differentiable Marching Tetrahedra variant with true gradients for sparse-view 3D human body reconstruction. More recently, \citet{maruani2023voromesh} proposed Voromesh, leveraging the regularity of Voronoi diagrams to optimize space tessellation for improved mesh quality. However, most existing differentiable methods treat grid optimization and SDF optimization as separate stages, potentially limiting the optimality of the final reconstruction.

\vspace{-1mm}
\section{Preliminaries}
\vspace{-1mm}

A set of $N$ 3D sites $ \{\site_i\}_{i=1}^N \subset \mathbb{R}^3$ implicitly represents a tessellation of the 3D space into Voronoi cells $\voronoicell(\site_i)$ defined as:  
\begin{equation*}
    \mathcal{V}(\site_i) = \{\point \in \mathbb{R}^3 \mid \|\point-\site_i\| < \|\crv{\point}-\site_j\|,\ \forall j \neq i\}.
\end{equation*}

The Voronoi diagram is also the dual of the Delaunay tetrahedralization $ \tetra = \{ \tetra_m \}_{m=1}^M $. 

Each tetrahedron $\tetra_m \subset \tetra$ with corresponding sites $\{\site_a, \site_b, \site_c, \site_d\}$ defines one Voronoi vertex $\vertex_m$ as its circumcenter, which is computed as
\begin{equation} \label{eq:voronoivertices}
\vertex_m
\, = \, \site_a
\, + \, \frac{
    \alpha\,(\mathbf{q}\times\mathbf{r})
  + \beta\,(\mathbf{r}\times\mathbf{p})
  + \gamma\,(\mathbf{p}\times\mathbf{q})
}{
    2\,\mathbf{p}\cdot(\mathbf{q}\times\mathbf{r})
}\,,
\end{equation}
where 
$\mathbf{p} = \site_b - \site_a$, 
$\mathbf{q} = \site_c - \site_a$, 
$\mathbf{r} = \site_d - \site_a$ 
are the local edges, 
and $\alpha,\beta,\gamma$ are their squared norms.  
The denominator is $2$ times the scalar triple product, i.e., $12\,\tetravol_\text{signed}$,  
with the tetrahedron volume $\tetravol = |\mathbf{p}\cdot(\mathbf{q}\times\mathbf{r})|/6$.

\mypar{Voronoi cell centroid.}
In general, a Voronoi diagram initialized with random sites produces an irregular discretization, where sites often form elongated tetrahedra and slivers\footnote{Flattened and elongated Voronoi cells}, which can cause significant issues during 3D mesh extraction. To improve the regularity of the tessellation, minimizing the Centroidal Voronoi Tessellation (CVT) energy is a widely used and robust strategy. A Voronoi diagram is said to be in a CVT configuration when all sites coincide with the centroids of their corresponding Voronoi cells. This configuration can be enforced using the Lloyd algorithm~\cite{lloyd1982algo}.

\vspace{-1mm}
\section{Proposed method} \label{sec:method}
\vspace{-1mm}

\begin{figure*}[t]
    \centering
    \includegraphics[width=\textwidth]{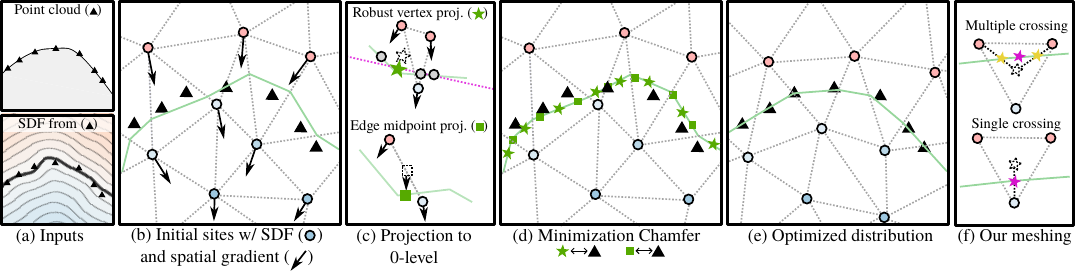}
    \vspace{-7mm}
    \caption{Overview of our proposed method. \crv{Starting from an unoriented 3D point cloud and an inferred SDF estimation from the point cloud (a)}, we initialize Voronoi sites (b), which yield a discretized SDF (zero-level shown in green). We then project zero-crossing Voronoi vertices (green stars) and edge midpoints (green squares) onto the SDF zero-level and minimize the Chamfer distance to the input point cloud (c–d). This optimization produces an improved site distribution and SDF representation (e), from which we extract the final mesh using our Voronoi-based meshing strategy (f), with Voronoi vertices shown in purple.}
    \vspace{-5mm}
    \label{fig:overview}
\end{figure*}

We address the problem of reconstructing an accurate, watertight 3D mesh from an unoriented point cloud (\cref{fig:overview}a). The surface is modeled as an implicit SDF discretized on a 3D grid (\cref{fig:overview}b), from which the mesh is extracted. Unlike prior work that optimizes only the SDF or the sampled site positions, we jointly optimize both the site positions $\{\site_i\}_{i=1}^N$ and their associated SDF values $\{\sdf_i\}_{i=1}^N$. Our approach comprises: (1) projecting 0-crossing Voronoi vertices $\{\vertex_j\}_{j=1}^M$ onto the SDF zero-level set via a robust projection scheme (\cref{sec:sdfprojection}); (2) regularizing the SDF-aware Voronoi diagram to satisfy the Centroidal Voronoi Tessellation (CVT) property (\cref{sec:regularization}); and (3) adaptively refining the discretization through error-driven site insertion (\cref{sec:upsampling}). The optimization is fully differentiable, facilitating integration with modern machine learning frameworks. \Cref{fig:overview} gives an overview of our proposed method. All diagrams are presented in 2D for simplicity.

\vspace{-1mm}
\subsection{Joint optimization} \label{sec:sdfprojection}
\vspace{-1mm}

The central idea of our method is to transform the conventional Voronoi diagram into a surface-aware Voronoi diagram, in which Voronoi vertices generated by tetrahedrons intersecting the SDF zero level are positioned close to the vertices of the final 3D mesh extracted by clipping the Voronoi cells. By formulating both the CVT regularization and the vertex projection onto the SDF zero level in a differentiable manner, we jointly optimize the SDF values and the discretization in an end-to-end framework, enabling loss functions to be applied directly to the vertices of the output 3D mesh.

In the remainder of this section, we restrict our attention to tetrahedrons that intersect the zero-level set of the SDF defined as $\tetra^{(0)}$. They are the elements that determine the vertices of the output 3D mesh. A tetrahedron $\tetra_j$, defined by the four sites $\{\site_a, \site_b, \site_c, \site_d\}$, is classified as intersecting if the SDF values at its summits satisfy
\begin{equation} \label{eq:zero-crossing}
\min(\sdf_a, \sdf_b, \sdf_c, \sdf_d) < 0 < \max(\sdf_a, \sdf_b, \sdf_c, \sdf_d)\,.
\end{equation}

For each such tetrahedron, we compute the corresponding Voronoi vertex $\vertex_j$ using \cref{eq:voronoivertices}, and subsequently project it onto the SDF zero-level. Likewise, the midpoints of all crossing edges are determined and projected onto the zero-level set. The data loss is then evaluated as the distance between these projected points and the input target points. By expressing the projection process in a differentiable form with respect to both the SDF values and the site positions, we enable joint optimization within a unified framework.

\mypar{Projection of 0-crossing Voronoi vertices.}

\Cref{fig:overview} (c) illustrates the projection of Voronoi vertices onto the zero-level set of the SDF. Each Voronoi vertex is defined as the circumcenter of its corresponding tetrahedron. However, this circumcenter can sometimes lie far from the tetrahedron itself, making straightforward barycentric interpolation of SDF values and gradients unreliable. Moreover, explicitly searching for the tetrahedron that contains a given Voronoi vertex is computationally prohibitive. To address these issues, we introduce an efficient and robust projection strategy that remains effective even under challenging tetrahedral configurations. Specifically, for each tetrahedron we fit a plane to the zero-level projections of its four summits and project the corresponding Voronoi vertex onto it.

For each $\tetra_i$ that crosses the zero level set, we first project its associated sites \({\site}_i'\) onto the zero-level set as
\begin{equation} \label{eq:projection-sites}
\site_i' \, = \, \site_i - \frac{ \nabla \sdf_i}{\|\nabla \sdf_i\| + \epsilon}\; \sdf_i\,.
\end{equation}
We then compute the centroid of the projected sites as
\begin{equation}
\bar{\site}'_j = \frac{1}{4} \sum_{i \in \tetra_j} \site'_i,
\end{equation}
We find the plane, defined by normal vector $\mathbf{n}_j$, which passes by the site centroid $\bar{\site}'_j$ and whose average distance to projected sites $\site_i'$ is minimum. Denoting the centered coordinates as $ \Delta \site'_i = \site'_i - \bar{\site}'_j$, this can be done by optimizing the following problem:
\begin{equation} \label{eq:minplane}
\arg\min_{\mathbf{n}_j} \ \frac{1}{4}\sum_{i \in \tetra_j} \Big(\!\underbrace{\frac{\mathbf{n}_j}{\|\mathbf{n}_j\|} \cdot \Delta \site'_i}_{\text{distance to plane}}\!\Big)^2.
\end{equation}
This normal vector \( \normal_j \), which defines the orientation of the fitted plane passing through the centroid \( \mathbf{c}_j \), corresponds to the first eigenvector\footnote{Eigenvector corresponding to the smallest eigenvalue.} of the centered site covariance matrix. Finally, we project the Voronoi vertex \( \vertex_j \) onto this fitted plane as
%
%

\begin{equation} \label{eq:projection-vertex-plane}
\vertex_j' \, = \, \vertex_j - \left[ (\vertex_j - \mathbf{c}_j) \cdot \normal_j \right] \, \normal_j \,.
\end{equation}
This projection ensures that the Voronoi vertex lies on the zero-level set of the SDF using only site-based information. See supplemental material for the full derivation (\cref{supp:robustplane}) and comparative experiments with using a naive Newton projection.

\mypar{Projection of 0-crossing edge midpoints.}
In addition to the Voronoi vertices, we compute the midpoints of tetrahedrons edges that cross the 0 level-set (\cref{fig:overview}c). These edge midpoints provide an additional constraint, ensuring that the Voronoi faces remain well aligned with both our SDF representation and the target point cloud. For each edge corresponding to a pair of adjacent sites \( (\site_i, \site_j) \) that crosses the zero-level set, we approximate the SDF value and the gradient vector of the midpoint $\mathbf{b}_{ij}$ as follows. 

\begin{align} \label{eq:midpoint-sdf}
\phi(\mathbf{b}_{ij}) \approx \tfrac{1}{2}(\phi_i + \phi_j), \,\, 
\nabla \phi(\mathbf{b}_{ij}) \approx \tfrac{1}{2}(\nabla \phi_i + \nabla \phi_j)
\end{align}
Finally, we project the edge midpoint \( \mathbf{b}_{ij} \) onto the zero-level set of the SDF using the Newton step method (\cref{eq:projection}).

\begin{equation} \label{eq:projection}
\mathbf{b}_{ij}' \, = \, \mathbf{b}_{ij} - \frac{ \nabla \phi(\mathbf{b}_{ij})}{\|\nabla \phi(\mathbf{b}_{ij})\| + \epsilon}\; \phi(\mathbf{b}_{ij})\,,
\end{equation}

\mypar{Spatial SDF gradient.}
To compute the spatial gradient of the SDF at each site, we assume a constant gradient within each tetrahedron, denoted by $\nabla \sdf^{(j)}$. Using a first-order approximation, the SDF at site $i \in \tetra_j$ is estimated as

\begin{equation}
\wh{\phi}_i \, \approx \, \bar{\phi}_j \, + \,  (\site_i - \bar{\site}_j)^\top \nabla\sdf^{(j)}. 
\end{equation}
The gradient $\nabla\sdf^{(j)}$ is then obtained by minimizing the mean squared difference between the actual and estimated SDF values over all sites:

\begin{equation} \label{eq:mingradsdf}
\arg\min_{\nabla\sdf^{(j)}} \ \frac{1}{4} \sum_{i \in \tetra_j} \big(\wh{\phi}_i - \phi_i\big)^2.
\end{equation}
As detailed in \cref{supp:derivgradsdf} inside the supplementary materials, this can be done efficiently by solving a linear system.

Then, the spatial gradient at site $\site_i$ is obtained as the volume-weighted average of the gradients of its incident tetrahedron:

\begin{equation} \label{eq:gradient-site}
\nabla \sdf_i \, = \, 
\frac{1}{\sum_{j \in \adjtetraset_\site} \tetravol_j} 
\sum_{j \in \adjtetraset_\site} \tetravol_j \nabla \sdf^{(j)},
\end{equation}
where $\adjtetraset_\site$ is the set of tetrahedrons adjacent to $\site$, $\tetravol_j$ is the volume of tetrahedron $j$, and $\nabla \sdf^{(j)}$ is its local gradient defined in \cref{eq:gradient-tetrahedron}.

\vspace{-1mm}
\subsection{Regularization} \label{sec:regularization}
\vspace{-1mm}

The data loss guiding our optimization is the Chamfer distance between the original point cloud and the set of reconstructed vertices and edge midpoints, defined as
\begin{equation} \label{eq:losschamfer}
\mathcal{L}_{\mathrm{CD}} \, = \, \min_{\mathrm{dist}}\,\big(\{ \|\targetpt - \mathbf{v}_i\|^2, \|\targetpt - \mathbf{b}_{ij}\|^2 \}\big)
\end{equation}
where \( \mathbf{v}_i \) are Voronoi vertices and \( \mathbf{b}_{ij} \) are edge midpoints. This loss encourages both the Voronoi vertices and edges to lie close to the input point cloud, ensuring that the extracted mesh accurately represents the input data. In addition to this main loss, we introduce several regularization losses to promote well-distributed sites and smooth SDF variation.

\mypar{CVT Regularization.}
A surface-aware Centroidal Voronoi Tessellation (CVT) loss is computed using vertices \textit{projected} onto the zero-level set (\cref{eq:projection-vertex-plane}). Vertices from tetrahedrons that do not intersect the zero-level set are left unchanged. During optimization, a site might deviate from this centroid, resulting in irregular Voronoi cells. To encourage regularity, we define a CVT loss that penalizes the Euclidean distance of each site $\site_i$ from its centroid $\centroid_i$ as
\begin{equation} \label{eq:losscvt}
\mathcal{L}_{\mathrm{CVT}} \, = \, \frac{1}{N} \sum_{i=1}^{N} \left\| 
\site_i - \centroid_i
\right\| \, .
\end{equation}

\mypar{Eikonal Regularization.} 
A fundamental property of SDFs is that they satisfy the Eikonal equation, which requires the gradient of the SDF to have unit norm at every point. This property is crucial for ensuring a consistent distance field and is extensively used in our multiple projection steps. To regularize the SDF gradient, we adapt the definition by \citet{wu2025sparse3D} and define our Eikonal loss as
\begin{equation} \label{eq:losseik}
\mathcal{L}_{\mathrm{Eik}} = \frac{1}{4M} \sum_{j=1}^{M} \sum_{i=1}^{4} \tetravol_j \big(\big\| \nabla \phi_{j,i} \big\|^2 - 1\big)^2 \,,
\end{equation}
where \( M \) is the number of tetrahedrons and \( \nabla \phi_{j,i} \) is the gradient of the SDF at summit \( i \) of tetrahedron \( j \). This encourages $\lVert \nabla \phi \rVert \approx 1$ throughout the domain, preserving the SDF property in a volume-weighted discrete form.

\mypar{Motion by mean curvature (MbMC) Regularization.} 
As shown by \citet{wu2025sparse3D}, the SDF can be regularized by incorporating a motion by mean curvature (MbMC) term, which encourages the SDF to evolve according to its mean curvature. This can be achieved by adding a term to the loss function that penalizes deviations from the mean curvature flow. For this they define a smeared-out Heaviside function \( H \) as
\begin{equation} \label{eq:heaviside}
H(\hat{\phi}) \, = \, 
\begin{cases}
0, & \hat{\phi} < -\epsilon_H \\
\frac{1}{2} + \frac{\hat{\phi}}{2\epsilon_H} + \frac{1}{2\pi} \sin\left( \frac{\pi \hat{\phi}}{\epsilon_H} \right), & -\epsilon_H \leq \hat{\phi} \leq \epsilon_H \\
1, & \hat{\phi} > \epsilon_H
\end{cases}
\end{equation}
where \( \epsilon_H \) is equal to the mean edge distance between sites, excluding the $5\%$ longer edges. By reusing the weight matrix \( \mathbf{W}_j \) defined in \cref{eq:gradient-tetrahedron} when solving for the tetrahedrons gradients, we can compute the mean curvature at each tetrahedron as
\begin{equation} \label{eq:mean-curvature}
\nabla \mathcal{H}^{(j)} = \mathbf{W}_j \begin{bmatrix}
H(\phi_{A}) - \bar{H}_j \\
H(\phi_{B}) - \bar{H}_j \\
H(\phi_{C}) - \bar{H}_j \\
H(\phi_{D}) - \bar{H}_j
\end{bmatrix},
\end{equation}
where \( \bar{H}_j \) is the mean of the Heaviside function values at the four sites of tetrahedron $\tetra_j$, similarly to the mean SDF value defined in \cref{eq:centering}. The MbMC loss is then defined as
\begin{equation} \label{eq:lossmbmc}
\mathcal{L}_{\mathrm{H}} = \frac{1}{M} \sum_{j=1}^{M} \tetravol_j \left\| \nabla \mathcal{H}^{(j)} \right\| \,,
\end{equation}
where \( M \) is the number of tetrahedrons and \( V_j \) is the volume of tetrahedron $\tetra_j$. 

\mypar{Loss.} 
The total loss of our optimization is  
\begin{equation} \label{eq:loss}
L = \mathcal{L}_{CD} + \lambda_{\mathrm{CVT}} \mathcal{L}_{CVT} + \lambda_{\mathrm{Eik}} \mathcal{L}_{Eik} + \lambda_{\mathrm{H}} \mathcal{L}_{H} \,,
\end{equation}
where \( \lambda_{\mathrm{CVT}} = 0.1, \lambda_{\mathrm{Eik}} = 0.02, \lambda_{\mathrm{H}} = 0.1 \) are the weights of the respective losses.

\vspace{-1mm}
\subsection{Upsampling} \label{sec:upsampling}
\vspace{-1mm}
One key aspect of our method is the ability to represent complex shapes with a limited number of Voronoi sites. Nevertheless, similar to prior approaches \cite{shen2021dmt,maruani2023voromesh}, our proposed method achieves higher mesh quality when more active sites are available (i.e., sites that have tetrahedrons crossing the zero-level set of the SDF) to improve the mesh quality. To this end, we employ an adaptive upsampling strategy that refines the reconstruction domain by inserting new sites according to local SDF features. The decision to insert additional sites is guided by a scoring function based on local spacing and curvature.

\mypar{Local Feature Computation.}
We compute local features for each Voronoi site \( \site_i \) to guide the upsampling process. The first feature is the local spacing \( \rho_i \), which measures the distance to the nearest neighbor site which is defined as
\begin{equation} \label{eq:localspacing}
\rho_{i} = \min_{j \in \mathcal{N}(i)} \|\site_i - \site_j\| \,,
\end{equation}
where \( \mathcal{N}(i) \) is the 1-ring neighborhood of site \( i \). This feature identifies regions where the actual density is low.

The second feature is the curvature proxy \( \kappa_{i} \), which measures the average variation of the SDF gradient in the neighborhood of site \( i \). This is computed as:
\begin{equation} \label{eq:curvatureproxy}
\kappa_{i} = \frac{\alpha_\kappa}{|\mathcal{N}(i)|} \sum_{j \in \mathcal{N}(i)} \left\| \wh{\nabla \phi}_i - \wh{\nabla \phi}_j \right\|^2 + (1-\alpha_\kappa),
\end{equation}
where \( \wh{\nabla \phi} = \nabla \phi / (\|\nabla \phi\| + \epsilon) \) is the unit SDF gradient and $\alpha_\kappa = 0.8$ is to give a non zero score on flat regions. This feature captures the local curvature of the SDF, which is essential for identifying regions that require more sites. 

\begin{figure}[t]
    \centering
    \includegraphics[width=\columnwidth]{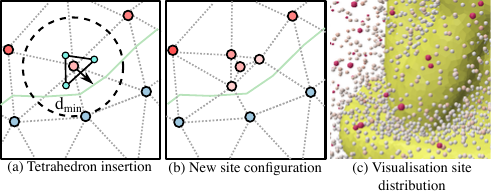}
    \caption{After selecting a site, we compute the minimal distance $\mathrm{d}_{\mathrm{min}}$ and insert a tetrahedron aligned with the SDF gradient (a). This preserves the local site connectivity (b). Our upsampling strategy results in a non-uniform site distribution (c).}
    \vspace{-5mm}
    \label{fig:upsampling-diag}
\end{figure}

\mypar{Score computation and candidate selection.}
We compute a score for each Voronoi site \( \site_i \) based on the local spacing and curvature features as follows: 
\begin{equation} \label{eq:score}
\score_i = \left( \frac{\rho_i}{\tilde{\rho}} \right) \left( \frac{\kappa_i}{\tilde{\kappa}} \right), \quad i \in \tetra^{(0)} ,
\end{equation}
where \( \tetra^{(0)} \) denotes the set of zero-crossing tetrahedron, \( \tilde{\rho} \) and \( \tilde{\kappa} \) are the median local spacing and curvature across all active sites, respectively. This score is designed to balance the uniform site spacing and curvature. Other sites producing a cell not crossing the zero-level receive a score of zero. We select sites candidates proportional to their score, by importance sampling proportionally to the score. A candidate site can be only selected once per upsampling iteration. 

\mypar{Scheduling.} We perform upsampling at regular intervals until $80\%$ of the optimization epochs, adding $10\%$ more sites at each step relative to the current number of sites. In our experiments, we allow a maximum of $10$ upsampling steps. To ensure a fair comparison with non-progressive upsampling methods, we cap the total number of sites.

\mypar{Tetrahedral Insertion.} 
One way to insert new sites is to randomly sample a point over a hemisphere centered at the candidate site and oriented along the spatial gradient of the SDF and based on its SDF distance. However, adding a single point can disturb the local connectivity between the Voronoi sites. Instead, we propose to insert a small tetrahedral structure around each selected site candidate. This approach allows us to maintain the local connectivity and regularity of our previous Voronoi tessellation. As shown in \cref{fig:upsampling-diag}, for each selected site \( \site_i \), we spawn 4 new sites forming a regular tetrahedron aligned with the spatial SDF gradient. Let \( \{ \mathbf{d}_k \}_{k=1}^4 \) be the canonical tetrahedral directions and \( \mathbf{T}_i \in \mathbb{R}^{3 \times 3} \) the local frame at \( \site_i \), then:

\begin{equation}
\site_{i,k}' = \site_i + \frac{\rho_i}{4} \mathbf{T}_i \mathbf{d}_k, \quad k = 1,\dots,4 \, ,
\end{equation}
where \( \rho_i \) is the local spacing defined in \cref{eq:localspacing} and \( \site_{i,k}' \) are the new sites to be inserted. We scale the regular tetrahedron using 1/4 of the local spacing to avoid adding new site which will cross the zero-level set. For each new site, we compute the SDF by using the SDF value at the original site \( \site_i \) and the spatial gradient \( \nabla \phi_i \) as follows:
\begin{equation}
\phi(\site_{i,k}') \approx \phi(\site_i) + \nabla \phi_i^\top (\site_{i,k}' - \site_i).
\end{equation}

\vspace{-1mm}
\subsection{Mesh extraction} \label{sec:meshing}
\vspace{-1mm}

The final vertex positions can be computed using either \cref{eq:projection-vertex-plane} or \cref{eq:projection} for each vertex belonging to a tetrahedron intersecting the zero-level set, based on the optimized site positions and SDF values. However, because our SDF model relies on a linear approximation, Voronoi vertices are not guaranteed to lie exactly on the zero-level set. To ensure a consistent and watertight mesh, we adopt a simpler interpolation procedure inspired by Marching Tetrahedra~\cite{shen2021dmt}.

In this method, we first identify the active tetrahedrons that are crossing the zero-level set. For each active tetrahedron, we compute their associated (unprojected) vertices and their SDF values via barycentric interpolation. We then identify the edge formed by an active vertex and its associated sites that are crossing the zero-level set. Finally we interpolate the vertex position along this edge at the zero-level set of the SDF. If a vertex results from multiple intersecting edges, we average all interpolated positions to determine its final location (\cref{fig:overview}f).

\vspace{-1mm}
\section{Results} \label{sec:results}
\vspace{-1mm}

We assess the effectiveness of our method in reconstructing detailed 3D meshes from unoriented point clouds using the publicly available Thingi32 dataset, a subset of the Thingi10K dataset \cite{zhou2016thingi10k} containing numerous high-resolution ground-truth meshes.

\vspace{-1mm}
\subsection{Experimental setup}
\vspace{-1mm}

In all experiments, we generate input point clouds by uniformly sampling $9.6$k points from each ground-truth 3D mesh. The Voronoi sites are initialized on grids of varying resolutions, with a small random perturbation of $0.005$ added to their positions to avoid coplanarity. Both the input point clouds and the initial site positions are then normalized to lie within the cube $[-1, 1]^3$.

\mypar{SDF Initialization.} For the SDF representation, we adopt the state-of-the-art neural network Hotspot \cite{zimo2025hotspot}, where each model is overfitted to the input point cloud. We use two versions of the trained model: (1) a fully converged model trained for $10$k iterations and (2) an unconverged model trained for $500$ iterations. The former is used to initialize a precise SDF representation, while the latter is used to evaluate our method's performance under an imprecise SDF representation. Note that we study shape ID 398259 separately in \cref{fig:ball2head}, as HotStop failed to produce a reliable SDF representation for this case.

\mypar{Baselines.} We compare our method against Voromesh~\cite{maruani2023voromesh} and Marching Tetrahedra (MTet)~\cite{shen2021dmt} as baselines. 
MTet is driven solely by the Chamfer distance computed from the extracted vertices. Voromesh minimizes a plane-based distance function and only optimizes the site positions. Which is why small extrusions can be seen on \Cref{fig:teaser} (c). In contrast, both MTet and our method jointly optimize the site positions and their associated SDF values. All methods are re-implemented in PyTorch~\cite{paszke2019pytorch} and initialized with the same sites. \crv{All methods use Adam optimizer~\cite{kingma2015adam} with a learning rate of $5 \times 10^{-4}$ and $\beta$ coefficients set to $(0.8, 0.99)$. All experiments are executed for
a fixed $1$k iterations.}

Unlike Voromesh, both MTet and our method require a tetrahedral representation of the Voronoi sites, obtained via Delaunay triangulation. We use gDel3D~\cite{cao2014gpudelaunay} to accelerate this computation on the GPU. The triangulation is recomputed from scratch at every iteration. Exploring incremental updates to the triangulation as the Voronoi sites evolve remains an interesting direction for future work.

\mypar{Centroid approximation.} Computing the exact centroid is time consuming. 
To accelerate the process, we use an approximate centroid computation. Instead of evaluating the exact centroid of each Voronoi cell, we simply average the positions of the projected and un-projected vertices associated with the cell. In practice, we did not observe any noticeable difference in the final results.

\mypar{Evaluation metrics.} We assess mesh quality using three metrics: Chamfer Distance squared (CD), F1 score, and Normal Consistency (NC). To compute these metrics, we sample $1\mathrm{M}$ points and apply face-to-point projection with FCPW~\cite{FCPW}. The Chamfer distance measures how closely the reconstructed mesh aligns with the input point cloud, while the F1 score evaluates the ability of the mesh to capture the original shape. We set the F1 threshold to $0.003$ to ensure sensitivity to fine details. For readability, all reported Chamfer distances are scaled by a factor of $10^5$. In the tables and figures, we use a green color to highlight the best metric and blue color for second best.

\vspace{-1mm}
\subsection{Analysis}
\vspace{-1mm}

\begin{figure}[t]
    \centering 
    \includegraphics[width=1\columnwidth]{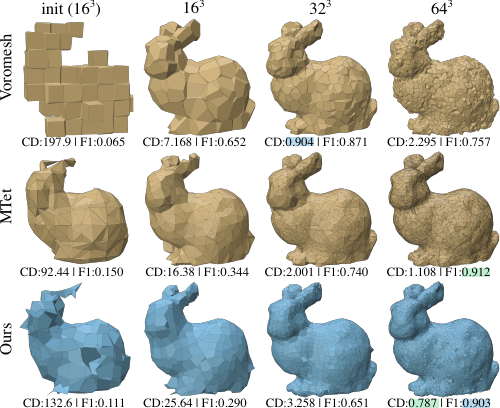}
    \caption{Optimized results for different resolution of sites with an accurate SDF representation.} 
    \vspace{-5mm}
    \label{fig:resolution}
\end{figure}
   
\mypar{Converged SDF example.} 
\Cref{fig:resolution} shows the results obtained by all methods for different initial site grid resolutions, and without upsampling. 
All methods, except Voromesh, remain stable across different grid resolutions for the initial site distribution. At higher resolutions, Voromesh receives too few projected target points per face, which leads to unstable face orientations. Although our $64^3$ configuration introduces some surface noise, the extracted mesh remains predominantly uniform thanks to our CVT regularization.
\begin{figure}[t]
    \centering
    \includegraphics[width=1\columnwidth]{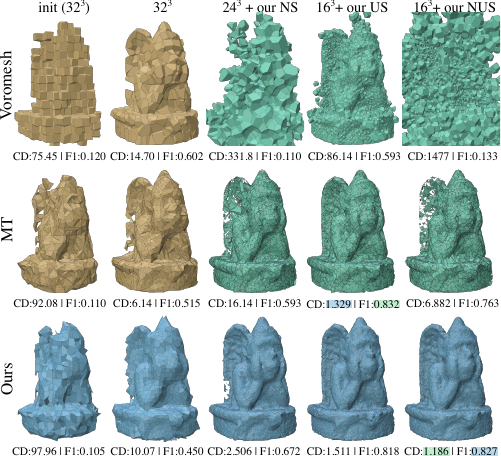}
    \caption{Results for different upsampling methods on an inaccurate SDF representation. Every optimization uses the same number of sites.}
    \vspace{-5mm}
    \label{fig:upsampling}
\end{figure}                            

\mypar{Unconverged SDF example.} \Cref{fig:upsampling} illustrates a more challenging scenario where the SDF representation is inaccurate. In this case, the initial, unoptimized representation exhibits large missing regions due to the SDF errors. As shown in the second column of \Cref{fig:upsampling}, both our method and MTet~\cite{shen2021dmt} are able to recover from this inaccurate SDF at a $32^3$ resolution. In contrast, Voromesh fails to recover, as it requires an accurate occupancy information for its final extraction.

\mypar{Non-uniform site distribution and upsampling.} \Cref{fig:upsampling}, other columns, present the results of three site placement strategies: (a) a naive approach, referred to as \textit{near sampling}, which places a proportion of the sites close to uniformly subsampled input point cloud, (b) our tetrahedral insertion upsampling strategy (\cref{sec:upsampling}), and (c) a combination of these two strategies. In this combination, we start with a uniform grid of $16^3$, add $16^3$ near sampling sites, and then apply our iterative upsampling strategy. In all cases, these alternative placement strategies result in a total of $32^3$ sites.

The near sampling strategy is effective only in our method, as we can efficiently move sites during optimization. In contrast, MTet suffers from this non-uniform site distribution, while Voromesh is unable to handle it due to instabilities in its target point projection. Our upsampling strategy improves performance by adaptively placing sites in regions with the highest reconstruction error proxy (\cref{eq:score}). Overall, combining near sampling with our upsampling strategy yields the best reconstruction results.

\begin{table}[t]
    \centering
    \setlength{\tabcolsep}{3pt}
    \renewcommand{\arraystretch}{1.2}
    \begin{small}
    \begin{tabular}{|l|c|c|c|c|c|c|}
        \hline
        \multirow{2}{*}{Method} & \multicolumn{3}{c|}{Converged} & \multicolumn{3}{c|}{Unconverged} \\
        \cline{2-7}
        & CD $\downarrow$ & F1 $\uparrow$ & NC $\uparrow$ & CD $\downarrow$ & F1 $\uparrow$ & NC $\uparrow$ \\
        \hline
        Voro $32^3$       & 10.96 & 0.628 & 0.922 & 14.70 & 0.602 &  0.902  \\
        MTet $32^3$       & 6.032 & 0.516 & 0.898 & 6.144 & 0.515 & 0.894 \\
        
        \cline{1-7}

        MTet U  & 2.539 & \cellcolor{secondcol} 0.682 & 0.916 & \cellcolor{secondcol} 2.443 &  \cellcolor{bestcol} 0.682 & 0.917 \\
        Ours U & 2.521 & 0.677 & 0.928 & 2.601 & 0.672 & 0.925 \\
        MTet N       & 7.400 & 0.636 & 0.873 & 16.14 & 0.593 & 0.838 \\
        Ours N   & \cellcolor{secondcol} 2.346 & 0.674 & \cellcolor{bestcol} 0.935 &  2.573 & 0.673 & \cellcolor{bestcol} 0.932 \\
        MTet NU & 5.051 & 0.660 & 0.893 & 10.02 & 0.619 & 0.864 \\
        Ours NU &  \cellcolor{bestcol} 2.326 & \cellcolor{bestcol} 0.684 & \cellcolor{secondcol} 0.930 & \cellcolor{bestcol} 2.276 & \cellcolor{secondcol} 0.680 & \cellcolor{secondcol} 0.930 \\
        
        \hline
    \end{tabular}
    \end{small}
    \caption{Average score over the 31 meshes of Thingi32 over the MTet and VoroMesh $32^3$ baseline versus our method. We include our near sample (N), upsampling (U) and combination (NU) approach for MTet and Ours for completeness.}
    \label{tab:thingi32}
\end{table}

\mypar{Average performance on Thingi32.} \Cref{tab:thingi32} reports the average performance of all methods on the Thingi32 dataset, excluding mesh ID 398259 due to its inaccurate SDF representation with HotSpot. Our method consistently outperforms both Voromesh and MTet with fixed resolutions, showing the benefits of our site placement strategies and optimization process. As shown in \cref{fig:upsampling}, our upsampling strategy also improves the performance of MTet, while our method remains robust to all adaptive site placement strategies thanks to the CVT formulation and its connection to projected vertices. On average, the combination of near sampling and upsampling yields the best overall results.
\crv{In terms of efficiency, our current implementation requires about $5$ minutes for optimization with $32^3$ sites on an RTX 3090 GPU—faster than MTet ($9$ minutes) but slower than Voromesh ($1$ minute). Leveraging approximations or more efficient optimization strategies could significantly reduce this computational overhead.}
\begin{figure}[t]
    \centering
    \includegraphics[width=1\columnwidth]{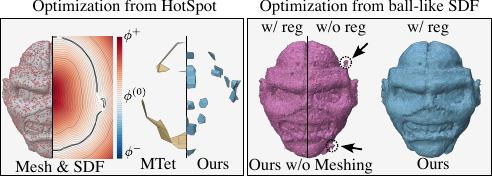}
    \caption{SDF interior/exterior issue causes mesh extraction issues for both MTet and Ours. Optimization from ball-like SDF works.}
    \vspace{-5mm}
    \label{fig:ball2head}
\end{figure}

\mypar{Simple SDF initialization.} \Cref{fig:ball2head} shows that the HotSpot SDF representation for mesh ID 398259 fails to correctly capture the interior–exterior relationship of the shape, due to the sparsity of sampled points. This poor initialization produces an extremely thin mesh volume, causing all methods that rely solely on zero-crossing information to fail. In contrast, since our method also optimizes the SDF representation, we can initialize from a simpler analytical SDF, such as a sphere, and still recover the target mesh. We note, however, that in this challenging case floating artifacts may appear when no regularization is applied to the SDF as visible in the purple mesh.

\vspace{-1mm}
\subsection{Ablation study}
\vspace{-1mm}
\begin{table}
    \centering
    \setlength{\tabcolsep}{6pt}
    \renewcommand{\arraystretch}{1.2}
    \begin{small}
    \begin{tabular}{|l|c|c|c|}
        \hline
        Method & CD $\downarrow$ & F1 $\uparrow$ & NC $\uparrow$ \\
        \hline
        Ours $32^3$            & 9.496 & 0.459 & 0.878 \\
        \hline
        w/o CVT Reg.     & 11.68 & 0.433 & 0.835 \\
        w/o Meshing      & 13.23 & 0.257 & 0.860 \\
        w/o Midpoint  & 11.63 & 0.441 & 0.851 \\
        \hline
    \end{tabular}
    \end{small}
    \caption{Ablation study of our method with $32^3$ grid resolution over the 31 shape of Thingi32.}
    \label{tab:ablation}
\end{table}
\Cref{tab:ablation} presents the ablation study for our different design choices using a $32^3$ site resolution and a converged SDF initialization. We observe that the CVT regularization (\cref{sec:regularization}) plays a crucial role in achieving high reconstruction quality, as also noted in earlier results. Our meshing strategy (\cref{sec:meshing}), compared to directly using the projected sites from \cref{eq:projection-sites}, yields better performance. Furthermore, projecting the midpoints (\cref{eq:midpoint-sdf}) is important for improving face orientation regularity. 

\vspace{-1mm}
\section{Conclusion and Future Work}
\vspace{-1mm}

We introduced DCCVT, a differentiable clipped centroidal Voronoi tessellation for 3D mesh reconstruction from point clouds. By projecting vertices onto the zero-level set while enforcing a CVT configuration, our method produces regular discretizations of optimized shapes. Furthermore, our framework supports different site placement strategies, enabling more accurate shape representations. Finally, our approach is capable of reconstructing meshes from various SDF representations, including unconverged or analytical initializations.
 
\crv{A promising future research direction is to combine DCCVT with deep learning frameworks to increase robustness when handling shapes with missing regions. Specifically, integrating visual foundation models could provide strong shape priors, allowing the system to plausibly fill gaps in the input point cloud and correct topological holes.}

\mypar{Acknowledgment}
\crv{
This work was supported by JSPS/KAKENHI JP23H03439 and AMED JP24wm0625404 at Kyushu University, and by the NSERC Discovery Grant RGPIN-2022-03182.
J. W. was supported by NSF and UC President's Postdoctoral Fellowships.
}

{
    \small
    \bibliographystyle{ieeenat_fullname}
    \bibliography{main}
}

\clearpage
\setcounter{page}{1}

\maketitlesupplementary \label{supp}
This supplemental document provides additional details on:  
\begin{itemize}
    \item our robust plane fitting method and its impact on the results (\cref{supp:robustplane});  
    \item the detailed derivation of \cref{eq:mingradsdf} (\cref{supp:derivgradsdf});  
    \item an analysis of how denser point clouds affect our optimization (\cref{supp:denserpc});  
    \item an ablation study on the effect of SDF regularization over the Thingi32 dataset (\cref{supp:sdfablation});  
    \item additional information and results (\cref{suppl:additonalresults});
\end{itemize}

\begin{figure}[t]
    \centering
    \includegraphics[width=1\columnwidth]{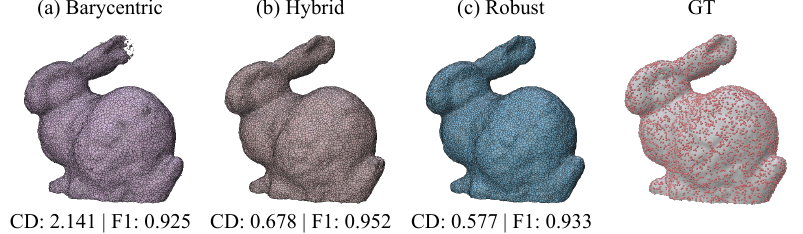}
    \caption{Visual comparison of barycentric (a), hybrid (b), and robust (c) projection methods against the ground truth (GT).}
    \label{fig:ablationproj}
\end{figure}
\section{Robust plane fitting} \label{supp:robustplane}

\Cref{eq:minplane} defines our minimization problem for robust plane fitting, which is used in the vertex projection step. In this section, we provide a more detailed description of how this optimization is carried out. We also present an ablation study demonstrating the effectiveness of our robust projection.

\mypar{Solution to \cref{eq:centering}.} The solution to this problem corresponds to the eigenvector with smallest eigenvalue of the covariance matrix
\begin{equation}
\mathbf{C}_j = \frac{1}{4} \sum_{i \in \tetra_j} \Delta \site'_i \, \Delta {\site'}_i^\top.
\end{equation}
and define the centered coordinates \( \Delta \site'_i = \site'_i - \bar{\site}'_j \). From these, we compute the covariance matrix \( \mathbf{C}_j \) of the projected sites:
\begin{equation}
\mathbf{C}_j = \frac{1}{4} \sum_{i \in \tetra_j} \Delta \site'_i \, \Delta {\site'}_i^\top.
\end{equation}
The plane normal \( \mathbf{n}_j \) is given by the eigenvector corresponding to the smallest eigenvalue of \( \mathbf{C}_j \), normalized as \( \mathbf{n}_j = \eigenvector_{\min} / \|\eigenvector_{\min}\|, \)
where \( \eigenvector_{\min} \) is the eigenvector of \( \mathbf{C}_j \) with the smallest eigenvalue.

\begin{proof}
To avoid the explicit normalization of $\mathbf{n}_j$ we first reformulate the problem as
\begin{equation}
\arg\min_{\mathbf{n}_j} \ \frac{1}{4}\sum_{i \in \tetra_j} \big(\mathbf{n}_j \cdot \Delta \site'_i\big)^2, \ \text{subject to } \|\mathbf{n}_j\|^2=1.
\end{equation}
We then model the added constraint with a Lagrangian formulation:
\begin{align}
\mathcal{L}(\mathbf{n}_j,\lambda) & \, = \, \frac{1}{4}\sum_{i \in \tetra_j} \big(\mathbf{n}_j \cdot \Delta \site'_i\big)^2 \ - \ \lambda\big(\|\mathbf{n}_j\|^2-1\big)\nonumber\\
& \, = \, \frac{1}{4}\sum_{i \in \tetra_j} \big(\mathbf{n}_j^\top \Delta \site'_i \big)\big(\Delta {\site'}_i^\top \mathbf{n}_j\big) \ - \ \lambda\big(\mathbf{n}_j^\top \mathbf{n}_j -1\big)\nonumber\\
& \, = \, \mathbf{n}_j^\top \Big(\frac{1}{4}\sum_{i \in \tetra_j} \Delta \site'_i \, \Delta {\site'}_i^\top \Big) \mathbf{n}_j \ - \ \lambda\big(\mathbf{n}_j^\top \mathbf{n}_j -1\big)\nonumber\\
& \, = \, \mathbf{n}_j^\top \mathbf{C}_j \mathbf{n}_j \ - \ \lambda\big(\mathbf{n}_j^\top \mathbf{n}_j -1\big).
\end{align}
Deriving this equation by $\mathbf{n}_j$ and setting the result to 0 yields
\begin{equation}
\mathbf{C}_j \mathbf{n}_j \, = \, \lambda \mathbf{n}_j, 
\end{equation}
hence $\mathbf{n}_j$ is an eigenvector of $\mathbf{C}_j$ associated to eigenvalue $\lambda$. The average distance to this plane corresponds to $
\mathbf{n}_j^\top \mathbf{C}_j \mathbf{n}_j = \lambda (\mathbf{n}_j^\top  \mathbf{n}_j) = \lambda$. Therefore we minimize the problem by finding the eigenvector with smallest eigenvalue. 
\end{proof}

\begin{table}[t]
    \centering
    \setlength{\tabcolsep}{6pt}
    \renewcommand{\arraystretch}{1.2}
    \begin{small}
    \begin{tabular}{|l|c|c|c|}
        \hline
        Method & CD $\downarrow$ & F1 $\uparrow$ & NC $\uparrow$ \\
        \hline
        Barycentric & 3.524 & 0.675 & 0.920 \\
        Hybrid & 2.418 & \cellcolor{bestcol}0.686 & 0.926 \\
        Robust (Ours) & \cellcolor{bestcol}2.326 & 0.684 & \cellcolor{bestcol}0.930 \\
        \hline
    \end{tabular}
    \end{small}
    \caption{Ablation study comparing our robust projection, hybrid and barycentric interpolation. Reported values are average errors on the Thingi32 dataset using our near sampling/upsampling approach with an unconverged SDF representation.}
    \label{tab:ablation_projection}
\end{table}
\mypar{Ablation study of our robust projection.} 
\Cref{fig:ablationproj} and \Cref{tab:ablation_projection} show the performance of our robust projection (c) compared to the barycentric (a) and hybrid (b) approaches, applied within our near sampling/upsampling setting. The barycentric method corresponds to projecting vertices using interpolated SDF and gradient values obtained from barycentric coordinates, followed by projection onto the zero-level set with \cref{eq:projection}. The hybrid method applies this barycentric projection only when the vertex lies inside its tetrahedron; otherwise, it falls back to the robust projection. We observe that the barycentric method generates the largest errors, as it is not robust to instabilities arising from extrapolation. The hybrid method alleviates this issue and significantly reduces the error, but abnormal spatial gradients can still occur, leading to slight instabilities during optimization. Applying our robust projection consistently in all cases provides the most stable and accurate results.

\section{Detailed derivation of \cref{eq:mingradsdf}} \label{supp:derivgradsdf}

\begin{proof}
Define the centroid and mean SDF value of tetrahedron $\tetra_j$ as
\begin{equation}
\bar{\site}_j = \tfrac{1}{4} \sum_{i \in \tetra_j} \mathbf{\site}_{i}, 
\quad
\bar{\phi}_j = \tfrac{1}{4} \sum_{i \in \tetra_j} \phi_{i}.
\end{equation}
We minimize the following problem
\begin{equation}\label{eq:sdf_approx}
\min_{\nabla\sdf^{(j)}} \ \frac{1}{4} \sum_{i \in \tetra_j} \big((\bar{\phi}_j \, + \,  (\site_i - \bar{\site}_j)^\top \nabla\sdf^{(j)}) - \phi_i\big)^2.
\end{equation}
To formulate this problem in a more compact manner, we define the vectors of centered coordinates and SDF values relative to the centroid and mean:
\begin{equation} \label{eq:centering}
\Delta \mathbf{S}_j = 
\begin{bmatrix}
\mathbf{\site}_{A} - \mathbf{\bar{\site}}_j \\
\mathbf{\site}_{B} - \mathbf{\bar{\site}}_j \\
\mathbf{\site}_{C} - \mathbf{\bar{\site}}_j \\
\mathbf{\site}_{D} - \mathbf{\bar{\site}}_j
\end{bmatrix},\quad
\Delta \Phi_j = 
\begin{bmatrix}
\phi_{A} - \bar{\sdf}_j \\
\phi_{B} - \bar{\sdf}_j \\
\phi_{C} - \bar{\sdf}_j \\
\phi_{D} - \bar{\sdf}_j
\end{bmatrix}.
\end{equation}
The problem of \cref{eq:sdf_approx} then becomes
\begin{align} \label{eq:mindetailed}
\arg\min_{\nabla\sdf^{(j)}} \ & \frac{1}{4} \big\|\Delta \mathbf{S}_j\nabla\sdf^{(j)} -  \Delta \Phi_j\big\|^2 \, = \nonumber\\
\arg\min_{\nabla\sdf^{(j)}} \ & \frac{1}{4} \big((\nabla\sdf^{(j)})^\top\mathbf{G}_j
\nabla\sdf^{(j)} - 2 (\nabla\sdf^{(j)})^\top\Delta\mathbf{S}_j^\top\Delta \Phi_j\big)
\end{align}
where $\mathbf{G}_j = \Delta\mathbf{S}_j^\top\Delta\mathbf{S}_j$ is the local Gram matrix. Deriving the above function with respect to $\nabla\sdf^{(j)}$ and setting the result to $0$ finally yields
\begin{equation}\label{eq:gradient-tetrahedron}
\nabla\sdf^{(j)} \, = \, \mathbf{G}_j^{-1} \Delta \mathbf{S}_j^\top \Delta \Phi_j = \mathbf{W}_j \Delta \Phi_j.
\end{equation}
\end{proof}

\section{Higher target point cloud density} \label{supp:denserpc}

\begin{figure*}[t]
    \centering
    \includegraphics[scale=0.95]{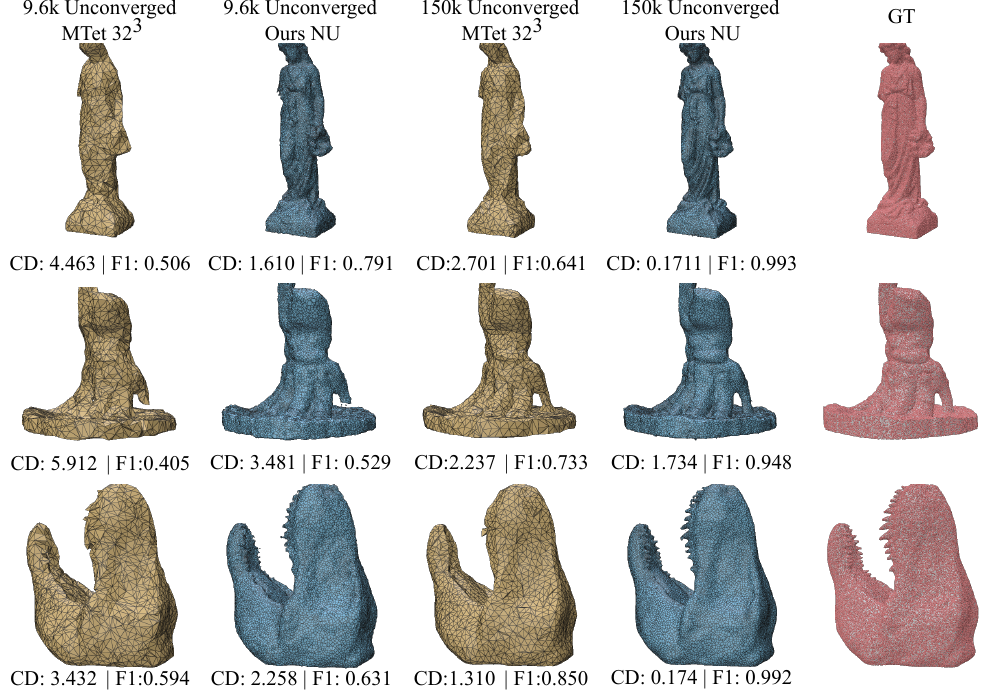}
    \caption{Comparison of MTet and our method (with near-sampling and upsampling, NU) under different target point cloud densities. Results are shown for $9.6k$ and $150k$ target points with an unconverged SDF. The ground truth (GT) is shown with $150k$ sampled points.}
    \vspace{-5mm}
    \label{fig:densepc}
\end{figure*}

\Cref{fig:densepc} compares MTet~\cite{shen2021dmt} and our method under different target point cloud densities with an unconverged SDF. Both methods benefit from increased point density in the target point cloud. However, our method with near sampling/upsampling consistently achieves the best results compared to the MTet baseline.

\section{Ablation of SDF regularization} \label{supp:sdfablation}

As shown in \cref{fig:ball2head}, SDF regularization is critical for removing small floating zero-crossing regions around the optimized mesh. While these artifacts are often small, they represent topological noise that degrades the cleanliness of the final reconstruction. 

Quantitatively, this trade-off is captured in \cref{tab:ablationSDF}. We observe that applying mild SDF regularization ($\lambda_\mathrm{Eik} = 0.02$) leads to a significant improvement in the Hausdorff Distance (HD), confirming the suppression of these spurious outliers. While this smoothing constraint results in a marginal decrease in surface precision—reflected by a slight rise in Chamfer Distance and lower F1 score compared to the unregularized baseline, it provides a necessary balance. It ensures the resulting SDF remains topologically consistent without the severe degradation seen at high regularization levels ($\lambda_\mathrm{Eik} = 2.00$).
\begin{table}[t]
    \centering
    \setlength{\tabcolsep}{6pt}
    \renewcommand{\arraystretch}{1.2}
    \begin{small}
    \begin{tabular}{|l|c|c|c|c|}
        \hline
        Method & CD $\downarrow$ & F1 $\uparrow$ & NC $\uparrow$ & HD $\downarrow$ \\
        \hline
        
        $\lambda_\mathrm{Eik} = 0.00, \lambda_\mathrm{H} = 0.00$ & 2.241 & 0.690 & 0.929 & 464.3\\
       
        $\lambda_\mathrm{Eik} = 0.02,\,\lambda_\mathrm{H}=0.10$ & 2.276 & 0.680 & 0.930 & 369.8\\
        
        $\lambda_\mathrm{Eik} = 2.00,\,\lambda_\mathrm{H}=10.0$ & 5.940	& 0.634 & 0.904 & 499.7\\
        \hline
    \end{tabular}
    \end{small}
    \caption{Ablation study on SDF regularization. We report average performance on the Thingi32 dataset for different values of the regularization weight $\lambda_{\text{sdf}}$.}
    \label{tab:ablationSDF}
\end{table}

\section{Additional information \& results} \label{suppl:additonalresults}

\crv{
 To ensure a rigorous and fair comparison, both MTet and DCCVT were initialized with the same initial sites and optimized using the Adam optimizer with the same hyperparameters ($LR=5 \times 10^{-4}$, $\beta=(0.8, 0.99)$) for 1000 iterations. A key distinction lies in the optimization process: while MTet is driven purely by Chamfer distance of extracted vertices, DCCVT incorporates CVT regularization to maintain discretization stability. Our experiments show that although both methods yield comparable results on regular grids, DCCVT exhibits superior robustness when initialized with non-uniform site distributions, such as near-sampling, where MTet tends to struggle. This robustness indicates that DCCVT's performance gains are intrinsic to its mathematical formulation rather than specific initialization advantages.}

\begin{figure*}[t]
    \centering
    \includegraphics[scale=0.92]{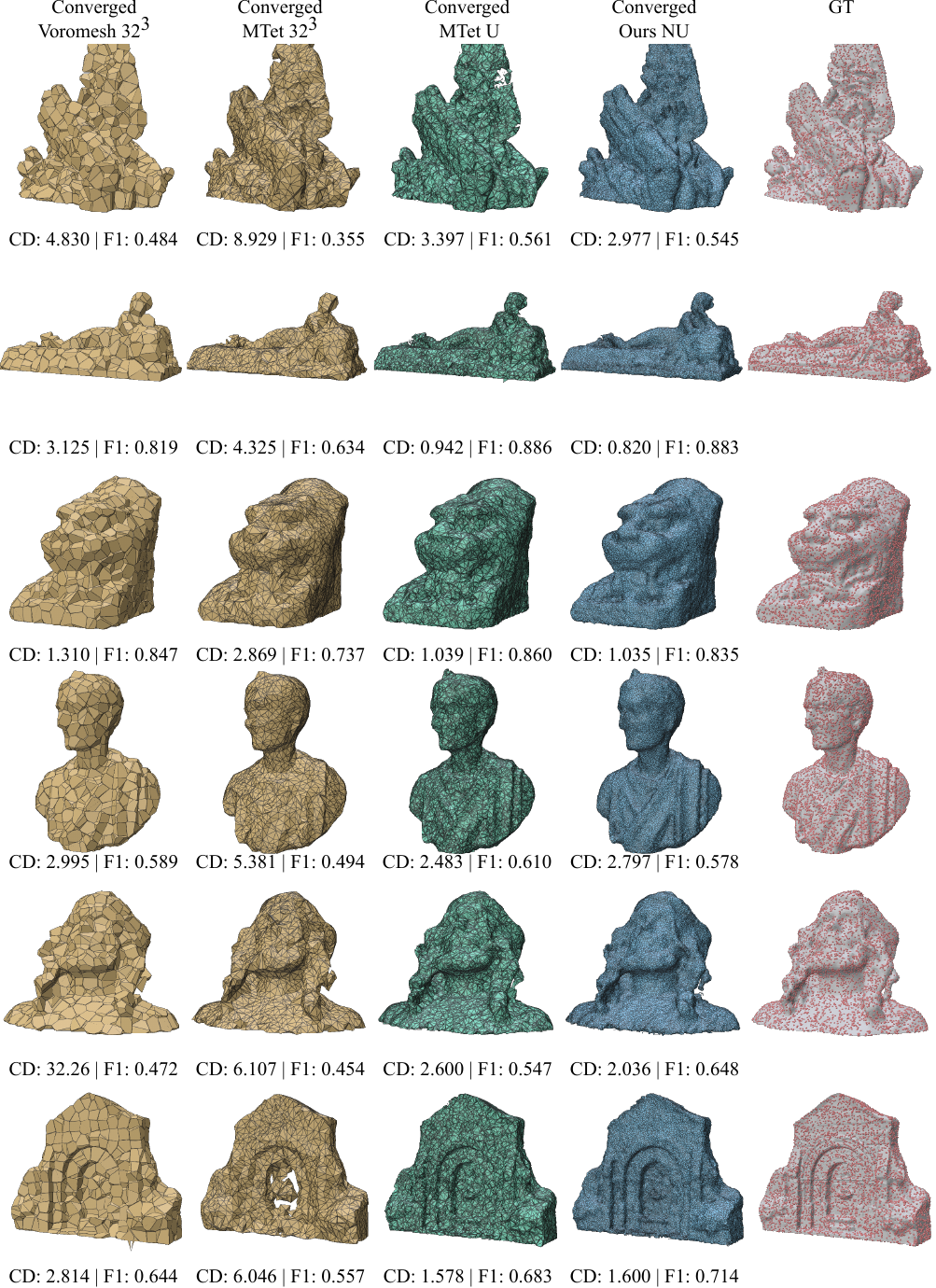}
   \caption{Comparison of Voromesh, MTet (with and without our upsampling), and our method with our near-sampling/upsampling) on converged SDFs. The ground truth (GT) is shown in the last column with $9600$ sampled points.}
    \vspace{-5mm}
    \label{fig:supp_conv_results}
\end{figure*}

\Cref{fig:supp_conv_results} shows additional results of our method with a converged SDF. For the Voromesh results, we applied a very small CVT regularization term to prevent co-planarity during mesh extraction. This increases the runtime of Voromesh, since a Delaunay triangulation must be recomputed at every iteration of the optimization process. We found this term to be necessary, particularly with unconverged SDF representations, where Voromesh frequently produces co-planar configurations in the final site distribution that are not supported by gDel3D~\cite{cao2014gpudelaunay}. As an alternative, CGAL could be used to perform the mesh extraction for Voromesh.

\begin{figure*}[t]
    \centering
    \includegraphics[scale=0.92]{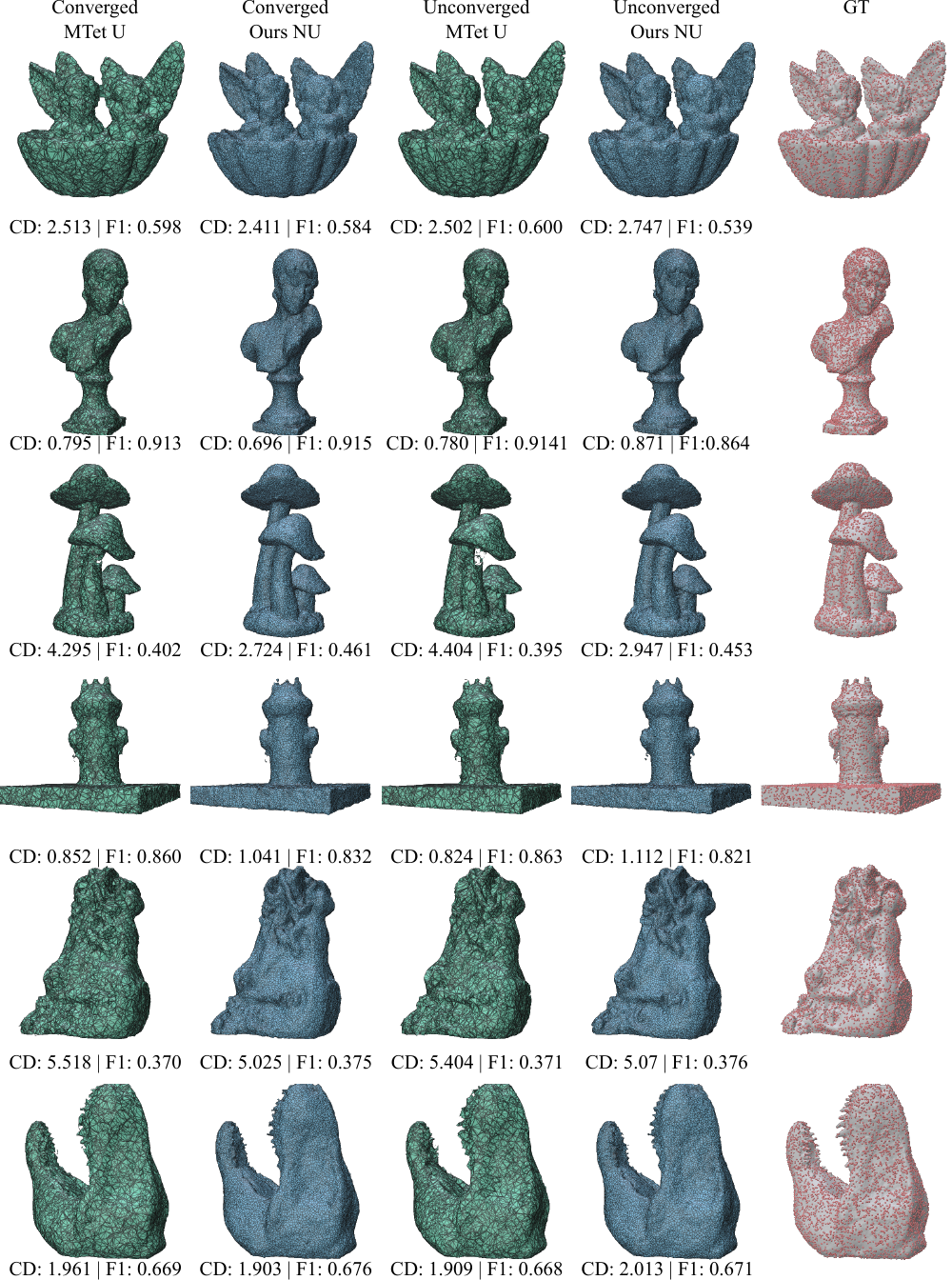}
    \caption{Comparison of MTet (with our upsampling) against ours (with our near-sampling/upsampling) in both converged and unconverged SDF case.}
    \vspace{-5mm}
    \label{fig:supp_conv_unconv_ours_MT_NU}
\end{figure*}

\cref{fig:supp_conv_unconv_ours_MT_NU} presents additional comparisons between our method and MTet under both converged and unconverged SDFs in the best-case of near sampling and upsampling setting. The results highlight the adaptivity and surface regularization of our approach, which generates significantly improved surface discretization.

\begin{figure}[h]
    \centering
    \includegraphics[scale=0.8]{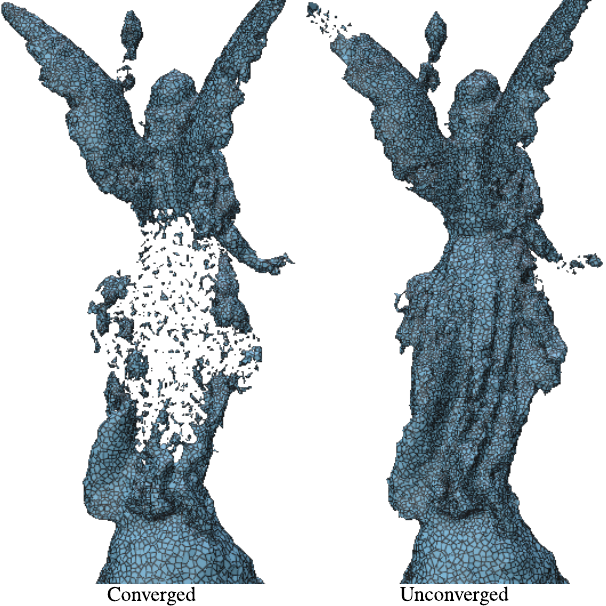}
    \caption{Hotspot converged produces a SDF that is very hard to recover from similarly to Mesh ID 398259 featured in \cref{fig:ball2head} }
    \vspace{-5mm}
    \label{fig:lucille}
\end{figure}

\begin{figure*}
    \centering
    \includegraphics[scale=0.75]{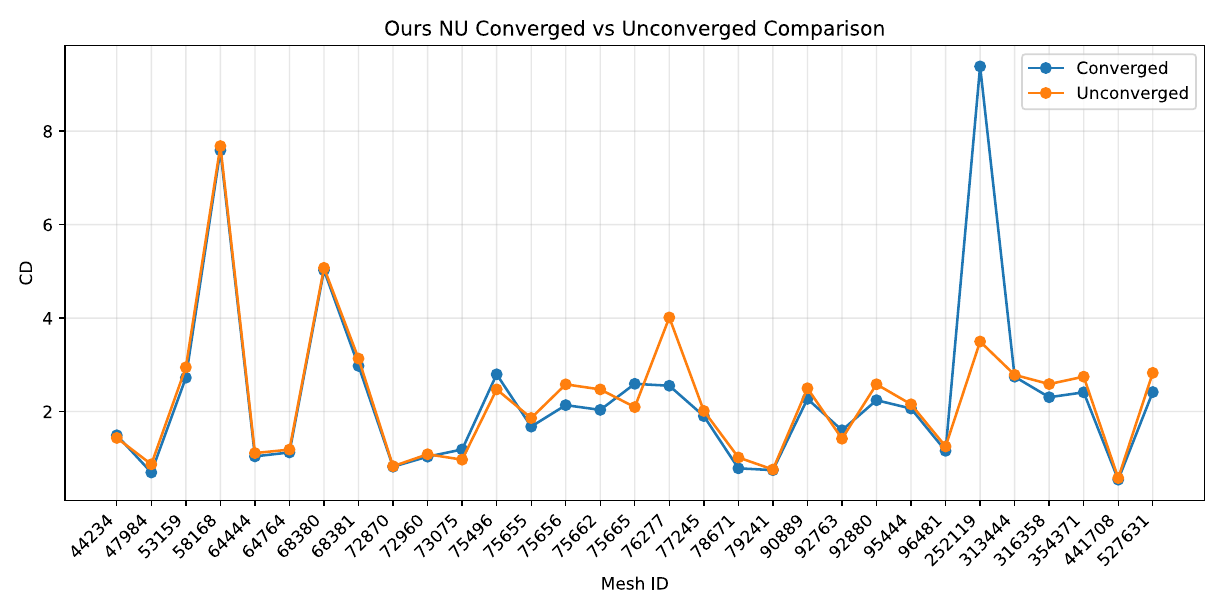}
    \caption{Per mesh Chamfer distance metric comparison between converged and unconverged Hotspot SDF state}
    \vspace{-5mm}
    \label{fig:linegraph}
\end{figure*}

\crv{\Cref{fig:linegraph} shows that while reconstruction quality is typically stable across SDF states, Mesh ID 252119 (\cref{fig:lucille}) suffers a catastrophic failure in the converged state. This mirrors the issue in Mesh ID 398259 (\cref{fig:ball2head}): the fully converged SDF provides a poor initialization that traps the optimization in a local minimum. The coarser, unconverged SDF avoids this pitfall, demonstrating that a less refined initialization can sometimes offer better regularization for the optimization loop.}

\end{document}